\documentclass[fdp,fleqn]{w-art}
\usepackage{times}
\usepackage{w-thm}

\usepackage{amssymb}
\newcommand{\bm}[1]{\mbox{\boldmath$#1$}}

\newcommand {\cD}{{\cal D}}
\newcommand {\cE}{{\cal E}}
\newcommand {\cF}{{\cal F}}
\newcommand {\cG}{{\cal G}}
\newcommand {\cH}{{\cal H}}

\newcommand {\cL}{{\cal L}}
\newcommand {\cM}{{\cal M}}
\newcommand {\cN}{{\cal N}}
\newcommand {\cO}{{\cal O}}

\newcommand {\cQ}{{\cal Q}}

\newcommand {\cV}{{\cal V}}
\newcommand {\cW}{{\cal W}}

\newcommand{\U}{\Upsilon}
\newcommand{\rd}{{\rm d}}
\newcommand{\ri}{{\rm i}}
\newcommand{\q}{\theta}

\newcommand{\ve}{\varepsilon}
\newcommand{\dalpha}{{\dot{\alpha}}}

\newcommand{\btheta}{{\bar\theta}}

\usepackage[]{graphicx}
\begin{document}
\DOIsuffix{theDOIsuffix}
\Volume{55}
\Issue{1}
\Month{01}
\Year{2007}
\pagespan{1}{}
\Receiveddate{xxx}
\Reviseddate{xxx}
\Accepteddate{xxx}
\Dateposted{xxx}



\title[Higher-derivative couplings]{Generating higher-derivative couplings in $\bm{\cN=2}$ supergravity}


\author[D. Butter]{Daniel Butter
  \footnote{Corresponding author\quad E-mail:~\textsf{daniel.butter@uwa.edu.au}, 
            Phone: +61\,8\,6488\,4224, 
            Fax: +61\,8\,6488\,7364.}}
\address{ School of Physics M013, The University of Western Australia\\
35 Stirling Highway, Crawley W.A. 6009, Australia}

\author[S. Kuzenko]{Sergei M. Kuzenko
	\footnote{E-mail:~\textsf{sergei.kuzenko@uwa.edu.au} }}
\begin{abstract}
Using a recently developed off-shell formulation for general 4D $\cN=2$
supergravity-matter systems, we propose a construction to generate higher
derivative couplings. We address here mainly the interactions of tensor and
vector multiplets, but the construction is quite general. For a certain
subclass of terms, the action is naturally written as an integral over
3/4 of the Grassmann coordinates of superspace.
\end{abstract}
\maketitle                   






\section{Introduction}

One of the advantages of superspace is that it makes supersymmetry manifest.
Any action which is invariant under arbitrary diffeomorphisms
of curved superspace automatically yields a locally supersymmetric component action.
In addition, it offers the possibility to construct general couplings from
Lagrangians of (essentially) arbitrary functional form.
For these reasons, higher derivative supersymmetric actions,
especially those coupled to supergravity, can be quite efficiently constructed
in superspace. Such actions have been of interest recently \cite{AHNT,dWKvZ}.

We review below a new class of higher derivative actions that was
constructed directly in superspace in our recent paper \cite{BK:HigherDeriv}.
The supergeometry which enables this construction is an off-shell formulation
for general $\cN=2$ supergravity-matter couplings in four dimensions \cite{KLRT-all},
which allows a curved-space extension of $\cN=2$ projective superspace \cite{(K)LR}.
For our purposes, the relevant detail is
that the required superspace is $\cM^{4|8} \times \mathbb CP^1$.
The curved supermanifold $\cM^{4|8}$ is parametrized by local coordinates
$z^M = (x^m, \theta_\imath^\mu, \bar\theta^\imath_{\dot \mu})$,
$\imath = \underline{1}, \underline{2}$. Its geometry is described by 
covariant derivatives
\begin{align}
\cD_A = E_A + \frac{1}{2} \Omega_A{}^{bc} M_{bc} + \Phi_A{}^{jk} J_{jk}
\end{align}
where $E_A = E_A{}^M \partial_M$ is the supervielbein; $M_{bc}$ and
$\Omega_A{}^{bc}$ are the Lorentz generators and superconnections respectively;
and $J_{jk}$ and $\Phi_A{}^{jk}$ are respectively the $\rm SU(2)$ generators and
superconnections. The auxiliary manifold $\mathbb CP^1$ 
is parametrized by an isotwistor $v^i \in \mathbb C^2 \setminus {0}$
defined modulo $v^i \sim c v^i$ for $c \in \mathbb C \setminus \{0\}$. All
superfields and operators are required to have fixed homogeneity in $v^i$.
For a more extensive discussion of the supergeometry,
we refer the reader to the original references \cite{KLRT-all}.

Our higher derivative construction is based on a duality
between two basic off-shell representations of $\cN=2$ supersymmetry --
the tensor multiplet and the vector multiplet. In curved superspace,
the tensor multiplet is described by its field strength $\cG^{ij}$, which is
a real isotriplet superfield, $(\cG^{ij})^* = \ve_{ik} \ve_{jl} \cG^{kl}$, constrained
to obey
$\cD_\alpha^{(i} \cG^{jk)} = \bar\cD_\dalpha^{(i} \cG^{jk)} = 0$.
These conditions are solved in terms of an unconstrained chiral
prepotential $\Psi$, $\bar\cD^\dalpha_i \Psi = 0$, as
\begin{align}\label{eq_GfromPsi}
\cG^{ij} = \frac{1}{4} (\cD^{ij} + 4 S^{ij}) \Psi + \frac{1}{4} (\bar \cD^{ij} + 4 \bar S^{ij}) \bar\Psi~.
\end{align}
The superfield $S^{ij}$ and its conjugate $\bar S_{ij}$ are components of the superspace
torsion tensor. The superspace torsion and the algebra of covariant derivatives
are given in \cite{KLRT-all}.

The vector multiplet, on the other hand, is described by its field strength $\cW$,
which is a chiral superfield, $\bar\cD^\dalpha_i \cW = 0$, obeying the Bianchi identity
\begin{align}\label{eq_BianchiW}
\Sigma^{ij} := \frac{1}{4} (\cD^{ij} + 4 S^{ij}) \cW = \frac{1}{4} (\bar \cD^{ij} + 4 \bar S^{ij}) \bar\cW~.
\end{align}
Such a superfield is called reduced chiral.
Comparing this equation to \eqref{eq_GfromPsi}, it is obvious that
the superfield $\Sigma^{ij}$ is a \emph{composite} tensor multiplet.
Moreover, it is clear that the tensor prepotential $\Psi$ is defined
only up to shifts $\delta \Psi = \ri \Lambda$ where $\Lambda$ is a reduced
chiral superfield. This is just the superfield version of the component gauge
transformation $\delta B_{mn} = 2 \partial_{[m} \Lambda_{n]}$ which leaves
the three-form field strength $H_{mnp} = 3\partial_{[m} B_{np]}$ invariant.

Within the projective superspace formulation of conformal supergravity \cite{KLRT-all},
both of these multiplets find a natural realization. The tensor multiplet
is described in terms of a real $\cO(2)$ multiplet
$\cG^{(2)}(v) := \cG^{ij} v_i v_j$ obeying
$\cD_\alpha^{(1)} \cG^{(2)} = \bar\cD_\dalpha^{(1)} \cG^{(2)} = 0$
where $\cD_\alpha^{(1)} := v_i \cD_\alpha^i$ and $\bar\cD_\dalpha^{(1)} := v_i \bar\cD_\dalpha^i$.
The vector multiplet can be described by a tropical prepotential $\cV$,
which is a real projective multiplet, $\cD_\alpha^{(1)} \cV = \bar\cD_\dalpha^{(1)} \cV = 0$,
of weight zero, $\cV(c v^i) = c \cV(v^i)$. The reduced chiral
superfield $\cW$ is related to $\cV$ by
\begin{gather}\label{eq_WfromV}
\cW  = \frac{1}{8\pi}  \oint_C v^k {\rm d}v_k
 \frac{u_i u_j}{(v^l u_l)^2} \Big( \bar \cD^{ij} + 4 \bar S^{ij}\Big) \cV~.
\end{gather}
The fixed isotwistor $u_i$ introduced here is subject only to
the condition that $v^k u_k \neq 0$ along the contour $C$
in $\mathbb CP^1$; the construction proves to be independent
of the choice of $u_i$.

The complementary nature of these properties allows us to generate
tensor multiplets from vector multiplets and vice-versa.
Consider a system of $n_V$ Abelian vector multiplets with field
strengths $\cW_I$, $I=1,\dots, n_V$. 
Given a holomorphic homogeneous function $F(\cW_I)$ of degree one
(to guarantee the correct super-Weyl transformation of $\mathbb G^{ij}$),
we can define a composite tensor multiplet
\begin{align}
{\mathbb G}^{ij}:=\frac{1}{4}\Big( \cD^{ij}+4S^{ij}\Big)F(\cW_I)
+ \frac{1}{4}\Big(  {\bar \cD}^{ij}+4{\bar S}^{ij}\Big){\bar F}({\bar \cW}_I)~.
\end{align}
This is a standard construction and a trivial application of \eqref{eq_GfromPsi}
using the chiral function $F$ as a composite tensor prepotential.

Conversely, we may take a system of $n_T$ tensor multiplets described by their field
strengths $\cG^{ij}_A$, with $A =1, \dots, n_T$, with
$\cG^{(2)}_A:= v_i v_j \cG^{ij}_A $ the corresponding $\cO(2)$ multiplets.
We may similarly construct a composite vector prepotential from any
function $\mathbb V(\cG^{(2)}_A)$ that is real and homogeneous of degree zero.
Applying \eqref{eq_WfromV} leads immediately to a composite field strength $\mathbb W$,
\begin{align}\label{eq_CompW}
\mathbb W  := \frac{1}{8\pi}  \oint_C v^k {\rm d}v_k
	\frac{u_i u_j}{(v^l u_l)^2} \Big( \bar \cD^{ij} + 4 \bar S^{ij}\Big) \mathbb V~.
\end{align}

These two observations together enable us to take any action involving tensor
and/or vector multiplets and convert it to a higher derivative action by
identifying one or both multiplets as composite. Iterating the procedure
leads to increasingly complex higher derivative interactions.
We first demonstrate how this procedure allows the construction of general
two-derivative interactions of tensor multiplets; then we extend to
two-derivative interactions of real $\cO(2n)$ multiplets; and then finally we extend
the argument to higher derivative theories.

\section{Self-couplings of tensor multiplets}

Consider the simplest interaction between vector and tensor multiplets:
the supersymmetric $BF$ coupling.
This action can be written either as
a chiral superspace integral involving the tensor prepotential $\Psi$
and vector field strength $\cW$,
\begin{align}\label{eq_PsiW}
S = \int \rd^4x\, \rd^4\theta\, \cE\, \Psi \cW + \textrm{c.c.} 
	= -\frac{1}{2} \int \rd^4x\, \varepsilon^{mnpq} \,B_{mn} F_{pq} + \cdots~,
\end{align}
or as a projective superspace integral involving the tensor
field strength $\cG^{(2)}$ and the vector prepotential $\cV$,
\begin{align}\label{eq_GV}
S = \frac{1}{2\pi} \oint_C v^i \rd v_i \int \rd^4x\, \rd^4\theta\, \rd^4\btheta
	\frac{E}{S^{(2)} \bar S^{(2)}} \,\, \cG^{(2)} \cV 
	= \frac{1}{3} \int \rd^4x\, \varepsilon^{mnpq} \,H_{mnp} A_{q} + \cdots~.
\end{align}
This action is topological and involves no propagating degrees of freedom;
however, by identifying one or both multiplets as composite, we may
construct nontrivial actions.

For example, the unique superconformal action for a single tensor multiplet,
known as the improved tensor multiplet action \cite{ImprovedTensor},
can be written in either form. In projective superspace, it is given
by \eqref{eq_GV} where $\cV$ is replaced by the composite vector prepotential
$\mathbb V = \ln(\cG^{(2)} / \ri \U^{(1)} \breve\U^{(1)})$.
The arctic multiplet $\U^{(1)}$ appearing here can be shown to be a pure gauge degree of freedom.
Equivalently, the action is given by \eqref{eq_PsiW} with the
corresponding replacement $\cW \rightarrow \mathbb W$.
This form is simplest to use in principle since the techniques
to evaluate the superspace integral \eqref{eq_PsiW} are well known:
we simply require $\mathbb W$. To construct it, we must evaluate
the contour integral in \eqref{eq_CompW}. The way this is usually done is to
make a choice for $u_i$ and then to evaluate the contour
explicitly, {\it e.g.} by representing $v^i = v^{\underline 1} (1, \zeta)$.
This breaks manifest $\rm SU(2)$ covariance. It is possible,
however, to keep $\rm SU(2)$ covariance along the way \cite{BK:HigherDeriv}.
One does this by first evaluating the spinor derivatives
on their argument and exploiting the contour integral to rewrite the integrand as
\begin{align}
\mathbb W
     &= \frac{1}{8\pi} \oint_C v^{i} \rd v_i \, \left(
     \frac{1}{3} \frac{\bar M}{\cG^{(2)}}
     - \frac{4}{9} \frac{\bar\chi^{(1)} \bar\chi^{(1)}}{(\cG^{(2)})^2}\right)
\end{align}
where ${\bar \chi}_{\dalpha}^{(1)} := v_i \bar\cD_{\dalpha k} \cG^{k i}$
and $\bar M := \left(\bar\cD_{jk} + 12 \bar S_{jk} \right) \cG^{jk}$.
This has the advantage that all dependence on the auxiliary isotwistor
$u_i$ has vanished. The remaining contour integral can be done
in an $\rm SU(2)$ covariant way \cite{BK:HigherDeriv}, with the result recast as
\begin{align}
\mathbb W
	= -\frac{\cG}{8} (\bar\cD_{ij} + 4 \bar S_{ij}) \left(\frac{\cG^{ij}}{\cG^2} \right)~,\qquad
\cG := \sqrt{\frac{1}{2} \cG^{ij} \cG_{ij}}~.
\end{align}
It is remarkable that this expression is chiral and obeys \eqref{eq_BianchiW}.
(An equivalent but less compact expression for $\mathbb W$,
obtained by brute force, was given in \cite{Muller86}.)

Superconformal actions involving several tensor multiplets can be
written in an analogous fashion. In projective superspace, they are
given by a general real weight-two projective Lagrangian
$\cL^{(2)}=\cL^{(2)}(\cG_A^{(2)})$ which is homogeneous of degree-one.
Such a Lagrangian can always be rewritten (though not uniquely) as
\begin{align}
\cL^{(2)} = \cG_A^{(2)} \mathbb V^A~,\qquad
\mathbb V^A = \mathbb V^A(\cG_B^{(2)})~.
\end{align}
The corresponding projective superspace action can then be converted
into a chiral action analogous to \eqref{eq_PsiW},
\begin{align}
S = \int \rd^4x\, \rd^4\theta\, \cE\, \Psi_A \mathbb W^A + \textrm{c.c.}
\end{align}
with $\mathbb W^A$ given by
\begin{gather}\label{eq_Wcomponent}
\mathbb W^A = \frac{1}{3} \cF^{A, B} \bar M_B
     + \frac{4}{9} \cF^{A, B, C}{}_{ij} \, \bar\chi^i_B \bar \chi^j_C
\end{gather}
where $\cF^{A,B}$ and $\cF^{A,B,C}{}_{ij}$ are functions of $\cG_A^{ij}$ given by
\begin{align}
\cF^{A, B} &:= \frac{1}{8\pi}  \oint_C v^i {\rm d} v_i \, \frac{\partial \mathbb V^A}{\partial \cG^{(2)}_B}~,
	\qquad
\cF^{A, B, C}{}_{ij} := \frac{\partial \cF^{A, B}}{\partial \cG_C^{ij}}~.
\end{align}
This construction appeared originally in flat superspace \cite{Siegel85}.
Its component form coupled to conformal supergravity appeared in \cite{deWS}.

\section{Adding $\cO(2n)$ multiplets}
We may extend the above procedure by considering $\cO(2n)$ multiplets.
As an example, consider a more general projective Lagrangian of the form
\begin{align}
\cL^{(2)} = \frac{\cQ^{(2n)}}{(\cG^{(2)})^{n-1}} = \cG^{(2)} \frac{\cQ^{(2n)}}{(\cG^{(2)})^{n}}
\end{align}
where $\cQ^{(2n)}$ is a real $\cO(2n)$ multiplet obeying
\begin{gather}
\cQ^{(2n)} = \cQ^{i_1 \cdots i_{2n}} v_{i_1} \cdots v_{i_{2n}}~,\quad 
(\cQ^{i_1 \cdots i_{2n}})^* = \cQ_{i_1 \cdots i_{2n}}~,\quad
\cD_\alpha^{(1)} \cQ^{(2n)} = \bar\cD_\dalpha^{(1)} \cQ^{(2n)} = 0~.
\end{gather}
The composite vector multiplet we construct from the above expression
has the prepotential $\mathbb V = \dfrac{\cQ^{(2n)}}{({\cG}^{(2)})^n}$.
As before, it is possible to evaluate the spinor derivatives in such a way
as to completely eliminate the auxiliary isotwistor $u_i$. The contour
integral \eqref{eq_CompW} can then be evaluated in an $\rm SU(2)$ covariant way and the
result recast as
\begin{align}
\mathbb W = -\frac{(2n)!}{2^{2n+2}\, (n+1)! (n-1)!}\,
     {\cG} \, (\bar\cD_{ij} + 4 \bar S_{ij}) \mathcal R^{ij}~,
\end{align}
where
\begin{align}
\mathcal R^{ij} = \frac{1}{\cG^{2n}}
     \left(\delta^{ij}_{kl} - \frac{1}{2 {\cG}^2} {\cG}^{ij} {\cG}_{kl} \right)
     \cQ^{kl \,i_1 \cdots i_{2n-2}}
      {\cG}_{i_1 i_2} \cdots {\cG}_{i_{2n-3} i_{2n-2}}  ~.
\end{align}

An interesting application of this result is for the case $\cQ^{(4)} = (\cH^{(2)})^2$
with the Lagrangian $\cL^{(2)} = \dfrac{(\cH^{(2)})^2}{{\cG}^{(2)}}$.
This is a curved-superspace version of that proposed in \cite{BSiegeldeWRV} to
describe the  classical universal hypermultiplet \cite{CFG}. Applying our
general formula leads to
\begin{align}
\cW = - \frac{\cG}{16} (\bar\cD_{ij} + 4 \bar S_{ij}) \mathcal R^{ij}~,\qquad
\mathcal R^{ij} = \frac{1}{ {\cG}^{4}} \left(\delta^{ij}_{kl} - \frac{1}{2 {\cG}^2} {\cG}^{ij} {\cG}_{kl} \right)
     \cH^{(kl} \cH^{m n)} {\cG}_{mn}~.
\end{align}
The construction of the component Lagrangian can then be carried out
by conventional means.

\section{Higher derivative couplings of tensor and vector multiplets}
We now turn to our main goal: the construction of higher derivative
actions. This can be done quite straightforwardly by iterating the
above procedure. Begin with a set of tensor multiplets $\cG_A^{ij}$
and Abelian vector multiplets $\cW_I$.
We construct a set of degree-zero functions $\mathbb V_{\hat I}$
of the tensor multiplets $\cG_A^{(2)}$, which lead to the composite vector multiplet
field strengths
\begin{align}\label{eq_CompWI}
\mathbb W_{\hat I} := \frac{1}{8\pi}  \oint_C v^k {\rm d}v_k
	\frac{u_i u_j}{(v^l u_l)^2} \Big( \bar \cD^{ij} + 4 \bar S^{ij}\Big)
		\mathbb V_{\hat I}(\cG_A^{(2)})~.
\end{align}
Using both sets of vector multiplets, we introduce a set of degree-one
holomorphic functions $F_{\hat A}(\cW_I, \mathbb W_{\hat I})$
which can be used to construct composite tensor multiplet field strengths
\begin{align}
{\mathbb G}_{\hat A}^{ij}:=\frac{1}{4}\Big( \cD^{ij}+4S^{ij}\Big)F_{\hat A}(\cW_I, \mathbb W_{\hat I})
+ \frac{1}{4}\Big(  {\bar \cD}^{ij}+4{\bar S}^{ij}\Big){\bar F}_{\hat A}({\bar \cW}_I, \bar{\mathbb W}_{\hat I})~.
\end{align}
Then we may introduce these composite tensor multiplets into
the functions $\mathbb V_{\hat I}$ in \eqref{eq_CompWI}, leading to
\begin{align}
\mathbb W_{\hat I} := \frac{1}{8\pi}  \oint_C v^k {\rm d}v_k
	\frac{u_i u_j}{(v^l u_l)^2} \Big( \bar \cD^{ij} + 4 \bar S^{ij}\Big)
		\mathbb V_{\hat I}(\cG_A^{(2)}, \mathbb G_{\hat A}^{(2)})~.
\end{align}
This new vector multiplet can be used to construct new composite tensor
multiplets and so on and so forth, with each iteration adding two spinor
derivatives (or one vector derivative) to the interaction.

This method of constructing higher derivative actions should be contrasted
with the more traditional way of generating higher derivative structures
using the chiral projection operator $\bar \Delta$, which is a curved
space generalization of $\bar D^4 = \bar D_{ij} \bar D^{ij} / 48$.
Given any scalar superfield $U(z)$ which is inert under super-Weyl transformations,
its descendant $\bar\Delta U$ is chiral and super-Weyl weight two.
Given a vector multiplet $\cW$ that is nowhere vanishing, 
we can then define the chiral scalar $\cW^{-2}  {\bar \Delta} U$ which is invariant
under the super-Weyl transformations. 
We can then construct ${\bar \cW}^{-2} \Delta(\cW^{-2}  {\bar \Delta} U)$,
and so on and so forth.\footnote{This 
is a generalization of the construction \cite{BKT}
 of rigid superconformal invariants containing $F^n$.}

Using these composite chiral operators, one may construct higher derivative actions 
involving chiral superspace actions. However, it is usually possible
to convert the chiral superspace action into an integral over the whole superspace
by eliminating one of the chiral projection operators. Schematically, if
$\cL_c = \Phi \bar\Delta U$ for some chiral superfield $\Phi$ and
a well-defined  local and gauge-invariant operator $U$, then
\begin{align}
\int \rd^4x\,\rd^4\theta\, \cE\, \Phi \bar\Delta U
     = \int \rd^4x\,\rd^4\theta\,\rd^4\bar\theta\, E\, \Phi \, U~.
\end{align}
Thus, higher derivative actions of this type are invariably most naturally
written as integrals over the entire superspace and are not intrinsically
chiral. This has important ramifications for perturbative
calculations, where non-renormalization theorems place strong restrictions
on intrinsic chiral Lagrangians.

The constructions we are considering are interesting partly because they
include higher derivative terms which \emph{cannot} be written as full superspace
integrals without introducing prepotentials. 
We take the example of a projective Lagrangian
$\cL^{(2)} = \mathbb G^{(2)} \mathbb V(\cG_A^{(2)})$
of several tensor multiplets $\cG_A^{(2)}$ and one composite
tensor multiplet
\begin{align}
\mathbb G^{(2)} = \frac{1}{4} ((\cD^{(1)})^2 + 4 S^{(2)}) F(\cW_I)
	+ \frac{1}{4} ((\bar\cD^{(1)})^2 + 4 \bar S^{(2)}) \bar F(\bar \cW_I)~.
\end{align}
The action can be rewritten
\begin{align}
S = \frac{1}{2\pi} \oint_C v^i \rd v_i \int \rd^4x\, \rd^4\q\, \rd^4\bar\q
	\, \frac{E}{\bar S^{(2)}} F(\cW_I) \mathbb V(\cG_A^{(2)}) + \textrm{c.c.}
\end{align}
which is a special case of a more general action
\begin{align}\label{eq_OmegaAction}
S = \frac{1}{2\pi} \oint_C v^i \rd v_i \int \rd^4x\, \rd^4\q\, \rd^4\bar\q
	\, \frac{E}{\bar S^{(2)}} \Omega(\cW_I, \cG_A^{(2)}) + \textrm{c.c.}~,\qquad
	\bar\cD_\dalpha^{(1)}\Omega = 0~.
\end{align}
The complex integrand $\Omega$ is required to be annihilated by only 1/4 of the
spinor derivatives. Such an action is the locally supersymmetric version
of
\begin{align}
S = -\frac{1}{8\pi} \oint_C v^k \rd v_k \int \rd^4x\,
	\frac{u_i u_j}{(v^l u_l)^2} \bar D^{ij} D^4 \Omega(\cW_I, \cG_A^{(2)}) + \textrm{c.c.}~,\qquad D^4 := \frac{1}{48} D^{ij} D_{ij}~.
\end{align}
In other words, it is an integral over 3/4 of the Grassmann coordinates
of superspace.\footnote{Special holomorphic 
three-derivative contributions to $\cN=2$ supersymmetric Yang-Mills  effective actions,
which are  given as an integral over 3/4 of superspace, 
have been discussed in \cite{Argyres:2003tg}.}
Because this class of higher derivative action cannot be written
as an integral over the full superspace without introducing gauge-dependent
prepotentials, an $\cN=2$ extension of the well known $\cN=1$ non-renormalization
theorems likely applies.

In order for the action \eqref{eq_OmegaAction} to be super-Weyl invariant,
$\Omega$ must be degree 1 in $\cW_I$ and degree zero in $\cG_A^{(2)}$. 
The simplest example is $\Omega = \cW \mathbb V$
where $\mathbb V = \ln (\cG^{(2)} / \ri \U^{(1)} \breve\U^{(1)})$. (In this
case, the arctic multiplets are again pure gauge degrees of freedom.)
A straightforward argument shows that the action can be rewritten
$\int \rd^4x\, \rd^4\q\, \cE\, W \mathbb W$
which is the supersymmetric generalization of
\begin{align}
\int \rd^4x\, e\, F^{pq} \mathbb F_{pq}
	= \frac{1}{4} \int \rd^4x\, e\, F^{pq} ( \ve_{kl} G_{ij} \cD_p G^{ik} \cD_q G^{jl}) / G^3 + \cdots~.
\end{align}
Thus the actions \eqref{eq_OmegaAction} describe three-derivative couplings
in the bosonic sector. Needless to say, the integrand $\Omega$ can be generalized to
include composite vector and tensor multiplets, which leads to four-derivative and
higher interactions.

For simplicity we have restricted our higher-derivative discussion to vector and tensor
multiplets. However, as emphasized in \cite{BK:HigherDeriv}, the composite
vector multiplets \eqref{eq_CompW} may be built out of not only tensor
multiplets, but also more general $\cO(2n)$ multiplets and even 
polar multiplets. The corresponding chiral superspace action
\eqref{eq_PsiW} can then be used to give a new and very general class
of higher-derivative supergravity-matter couplings.

\begin{acknowledgement}
This work is supported in part by the Australian Research Council and by
a UWA Research Development Award.
\end{acknowledgement}

\end{document}